\documentclass[showpacs,prl,aps,10pt,superscriptaddress,altaffilletter,floatfix,twocolumn]{revtex4-1}
\usepackage{graphicx}
\usepackage{hyperref}
\usepackage{amsmath}
\usepackage{amssymb}
\usepackage{color}
\usepackage{xspace}
\usepackage{dcolumn}
\newcolumntype{C}[1]{>{\centering\let\newline\\\arraybackslash\hspace{0pt}}m{#1}}

\RequirePackage{lineno}

\begin{document}

\newcommand{\isot}[2]{$^{#2}$#1}
\newcommand{\isotbold}[2]{$^{\boldsymbol{#2}}$#1}
\newcommand{\xeiso}{\isot{Xe}{136}\xspace}
\newcommand{\thsrc}{\isot{Th}{228}\xspace}
\newcommand{\cosrc}{\isot{Co}{60}\xspace}
\newcommand{\rasrc}{\isot{Ra}{226}\xspace}
\newcommand{\cssrc}{\isot{Cs}{137}\xspace}
\newcommand{\betascale}  {$\beta$-scale}
\newcommand{\kevkgyr}  {keV$^{-1}$ kg$^{-1}$ yr$^{-1}$}
\newcommand{\nonubb}  {$0\nu \beta\!\beta$\xspace}
\newcommand{\nonubbbf}  {$\boldsymbol{0\nu \beta\!\beta}$\xspace}
\newcommand{\twonubb} {$2\nu \beta\!\beta$\xspace}
\newcommand{\bb} {$\beta\!\beta$\xspace}
\newcommand{\vadc} {ADC$_\text{V}$}
\newcommand{\uadc} {ADC$_\text{U}$}
\newcommand{\mus} {\textmu{}s}
\newcommand{\chisq} {$\chi^2$}
\newcommand{\mum} {\textmu{}m}
\newcommand{\checkit}[1]{{\color{red}#1}}
\newcommand{\RunTwoA}{Run 2a}
\newcommand{\RunTwo}{Run 2}
\newcommand{\RunTwoBC}{Runs 2b and 2c}
\newcommand{\SP}[1]{\textsuperscript{#1}}
\newcommand{\SB}[1]{\textsubscript{#1}}
\newcommand{\SPSB}[2]{\rlap{\textsuperscript{#1}}\SB{#2}}
\newcommand{\pmasy}[3]{#1\SPSB{$+$#2}{$-$#3}}
\newcommand{\matel}{$M^{2\nu}$}
\newcommand{\psfac}{$G^{2\nu}$}
\newcommand{\tbeta}{T$_{1/2}^{0\nu\beta\beta}$}
\newcommand{\exolimit}[1][true]{\pmasy{2.6}{1.8}{2.1}$ \cdot 10^{25}$}
\newcommand{\exomeasurement}{\tbeta{}= \exolimit{}~yr}
\newcommand{\U}{\text{U}}
\newcommand{\V}{\text{V}}
\newcommand{\X}{\text{X}}
\newcommand{\Y}{\text{Y}}
\newcommand{\Z}{\text{Z}}
\newcommand{\bqcm}{${\rm Bq~m}^{-3}$}
\newcommand{\nonunorm}{N_{{\rm Err, } 0\nu\beta\beta}}
\newcommand{\nonunum}{n_{0\nu\beta\beta}}
\newcommand{\cussim}[1]{$\sim$#1}
\newcommand{\halflife}[1]{$#1\cdot10^{25}$~yr}
\newcommand{\numspec}[3]{$N_{^{#2}\mathrm{#1}}=#3$}
\newcommand{\TD}[1]{\textcolor{red}{#1}}
\newcommand{\PI}{Phase~I\xspace}
\newcommand{\PII}{Phase~II\xspace}
\newcommand{\Rn}{radon\xspace}

\title{Search for Neutrinoless Double-Beta Decay with the Upgraded EXO-200 Detector}

\author{J.B.~Albert}\affiliation{Physics Department and CEEM, Indiana University, Bloomington, Indiana 47405, USA}
\author{G.~Anton}\affiliation{Erlangen Centre for Astroparticle Physics (ECAP), Friedrich-Alexander University Erlangen-N\"urnberg, Erlangen 91058, Germany}
\author{I.~Badhrees}\altaffiliation{Permanent position with King Abdulaziz City for Science and Technology, Riyadh, Saudi Arabia}\affiliation{Physics Department, Carleton University, Ottawa, Ontario K1S 5B6, Canada}
\author{P.S.~Barbeau}\affiliation{Department of Physics, Duke University, and Triangle Universities Nuclear Laboratory (TUNL), Durham, North Carolina 27708, USA}
\author{R.~Bayerlein}\affiliation{Erlangen Centre for Astroparticle Physics (ECAP), Friedrich-Alexander University Erlangen-N\"urnberg, Erlangen 91058, Germany}
\author{D.~Beck}\affiliation{Physics Department, University of Illinois, Urbana-Champaign, Illinois 61801, USA}
\author{V.~Belov}\affiliation{Institute for Theoretical and Experimental Physics, Moscow, Russia}
\author{M.~Breidenbach}\affiliation{SLAC National Accelerator Laboratory, Menlo Park, California 94025, USA}
\author{T.~Brunner}\affiliation{Physics Department, McGill University, Montreal, Quebec, Canada}\affiliation{TRIUMF, Vancouver, British Columbia V6T 2A3, Canada}
\author{G.F.~Cao}\affiliation{Institute of High Energy Physics, Beijing, China}
\author{W.R.~Cen}\affiliation{Institute of High Energy Physics, Beijing, China}
\author{C.~Chambers}\affiliation{Physics Department, Colorado State University, Fort Collins, Colorado 80523, USA}
\author{B.~Cleveland}\affiliation{Department of Physics, Laurentian University, Sudbury, Ontario P3E 2C6, Canada}\affiliation{SNOLAB, Sudbury, Ontario P3Y 1N2, Canada}
\author{M.~Coon}\affiliation{Physics Department, University of Illinois, Urbana-Champaign, Illinois 61801, USA}
\author{A.~Craycraft}\affiliation{Physics Department, Colorado State University, Fort Collins, Colorado 80523, USA}
\author{W.~Cree}\affiliation{Physics Department, Carleton University, Ottawa, Ontario K1S 5B6, Canada}
\author{T.~Daniels}\affiliation{SLAC National Accelerator Laboratory, Menlo Park, California 94025, USA}
\author{M.~Danilov}\altaffiliation{Now at P.N.Lebedev Physical Institute of the Russian Academy of Sciences, Moscow, Russia}\affiliation{Institute for Theoretical and Experimental Physics, Moscow, Russia}
\author{S.J.~Daugherty}\affiliation{Physics Department and CEEM, Indiana University, Bloomington, Indiana 47405, USA}
\author{J.~Daughhetee}\affiliation{Department of Physics, University of South Dakota, Vermillion, South Dakota 57069, USA}
\author{J.~Davis}\affiliation{SLAC National Accelerator Laboratory, Menlo Park, California 94025, USA}
\author{S.~Delaquis}\affiliation{SLAC National Accelerator Laboratory, Menlo Park, California 94025, USA}
\author{A.~Der~Mesrobian-Kabakian}\affiliation{Department of Physics, Laurentian University, Sudbury, Ontario P3E 2C6, Canada}
\author{R.~DeVoe}\affiliation{Physics Department, Stanford University, Stanford, California 94305, USA}
\author{T.~Didberidze}\altaffiliation{Now University of Idaho, Moscow, Idaho, USA}\affiliation{Department of Physics and Astronomy, University of Alabama, Tuscaloosa, Alabama 35487, USA}
\author{J.~Dilling}\affiliation{TRIUMF, Vancouver, British Columbia V6T 2A3, Canada}
\author{A.~Dolgolenko}\affiliation{Institute for Theoretical and Experimental Physics, Moscow, Russia}
\author{M.J.~Dolinski}\affiliation{Department of Physics, Drexel University, Philadelphia, Pennsylvania 19104, USA}
\author{W.~Fairbank Jr.}\affiliation{Physics Department, Colorado State University, Fort Collins, Colorado 80523, USA}
\author{J.~Farine}\affiliation{Department of Physics, Laurentian University, Sudbury, Ontario P3E 2C6, Canada}
\author{S.~Feyzbakhsh}\affiliation{Amherst Center for Fundamental Interactions and Physics Department, University of Massachusetts, Amherst, MA 01003, USA}
\author{P.~Fierlinger}\affiliation{Technische Universit\"at M\"unchen, Physikdepartment and Excellence Cluster Universe, Garching 80805, Germany}
\author{D.~Fudenberg}\affiliation{Physics Department, Stanford University, Stanford, California 94305, USA}
\author{R.~Gornea}\affiliation{Physics Department, Carleton University, Ottawa, Ontario K1S 5B6, Canada}\affiliation{TRIUMF, Vancouver, British Columbia V6T 2A3, Canada}
\author{K.~Graham}\affiliation{Physics Department, Carleton University, Ottawa, Ontario K1S 5B6, Canada}
\author{G.~Gratta}\affiliation{Physics Department, Stanford University, Stanford, California 94305, USA}
\author{C.~Hall}\affiliation{Physics Department, University of Maryland, College Park, Maryland 20742, USA}
\author{E.V.~Hansen}\affiliation{Department of Physics, Drexel University, Philadelphia, Pennsylvania 19104, USA}
\author{J.~Hoessl}\affiliation{Erlangen Centre for Astroparticle Physics (ECAP), Friedrich-Alexander University Erlangen-N\"urnberg, Erlangen 91058, Germany}
\author{P.~Hufschmidt}\affiliation{Erlangen Centre for Astroparticle Physics (ECAP), Friedrich-Alexander University Erlangen-N\"urnberg, Erlangen 91058, Germany}
\author{M.~Hughes}\affiliation{Department of Physics and Astronomy, University of Alabama, Tuscaloosa, Alabama 35487, USA}
\author{A.~Jamil}\affiliation{Erlangen Centre for Astroparticle Physics (ECAP), Friedrich-Alexander University Erlangen-N\"urnberg, Erlangen 91058, Germany}\affiliation{Physics Department, Stanford University, Stanford, California 94305, USA}
\author{M.J.~Jewell}\affiliation{Physics Department, Stanford University, Stanford, California 94305, USA}
\author{A.~Johnson}\affiliation{SLAC National Accelerator Laboratory, Menlo Park, California 94025, USA}
\author{S.~Johnston}\altaffiliation{Now at Argonne National Laboratory, Argonne, Illinois, USA}\affiliation{Amherst Center for Fundamental Interactions and Physics Department, University of Massachusetts, Amherst, MA 01003, USA}
\author{A.~Karelin}\affiliation{Institute for Theoretical and Experimental Physics, Moscow, Russia}
\author{L.J.~Kaufman}\altaffiliation{Now at SLAC National Accelerator Laboratory, Menlo Park, California, USA}\affiliation{Physics Department and CEEM, Indiana University, Bloomington, Indiana 47405, USA}
\author{T.~Koffas}\affiliation{Physics Department, Carleton University, Ottawa, Ontario K1S 5B6, Canada}
\author{S.~Kravitz}\affiliation{Physics Department, Stanford University, Stanford, California 94305, USA}
\author{R.~Kr\"{u}cken}\affiliation{TRIUMF, Vancouver, British Columbia V6T 2A3, Canada}
\author{A.~Kuchenkov}\affiliation{Institute for Theoretical and Experimental Physics, Moscow, Russia}
\author{K.S.~Kumar}\affiliation{Department of Physics and Astronomy, Stony Brook University, SUNY, Stony Brook, New York 11794, USA}
\author{Y.~Lan}\affiliation{TRIUMF, Vancouver, British Columbia V6T 2A3, Canada}
\author{D.S.~Leonard}\affiliation{IBS Center for Underground Physics, Daejeon 34047, Korea}
\author{G.S.~Li}\affiliation{Physics Department, Stanford University, Stanford, California 94305, USA}
\author{S.~Li}\affiliation{Physics Department, University of Illinois, Urbana-Champaign, Illinois 61801, USA}
\author{C.~Licciardi}\email[Corresponding author: ]{licciard@triumf.ca}\affiliation{Physics Department, Carleton University, Ottawa, Ontario K1S 5B6, Canada}
\author{Y.H.~Lin}\affiliation{Department of Physics, Drexel University, Philadelphia, Pennsylvania 19104, USA}
\author{R.~MacLellan}\affiliation{Department of Physics, University of South Dakota, Vermillion, South Dakota 57069, USA}
\author{T.~Michel}\affiliation{Erlangen Centre for Astroparticle Physics (ECAP), Friedrich-Alexander University Erlangen-N\"urnberg, Erlangen 91058, Germany}
\author{B.~Mong}\affiliation{SLAC National Accelerator Laboratory, Menlo Park, California 94025, USA}
\author{D.~Moore}\affiliation{Department of Physics, Yale University, New Haven, Connecticut 06511, USA}
\author{K.~Murray}\affiliation{Physics Department, McGill University, Montreal, Quebec, Canada}
\author{R.~Nelson}\affiliation{Waste Isolation Pilot Plant, Carlsbad, New Mexico 88220, USA}
\author{O.~Njoya}\affiliation{Department of Physics and Astronomy, Stony Brook University, SUNY, Stony Brook, New York 11794, USA}
\author{A.~Odian}\affiliation{SLAC National Accelerator Laboratory, Menlo Park, California 94025, USA}
\author{I.~Ostrovskiy}\affiliation{Department of Physics and Astronomy, University of Alabama, Tuscaloosa, Alabama 35487, USA}
\author{A.~Piepke}\affiliation{Department of Physics and Astronomy, University of Alabama, Tuscaloosa, Alabama 35487, USA}
\author{A.~Pocar}\affiliation{Amherst Center for Fundamental Interactions and Physics Department, University of Massachusetts, Amherst, MA 01003, USA}
\author{F.~Reti\`{e}re}\affiliation{TRIUMF, Vancouver, British Columbia V6T 2A3, Canada}
\author{P.C.~Rowson}\affiliation{SLAC National Accelerator Laboratory, Menlo Park, California 94025, USA}
\author{J.J.~Russell}\affiliation{SLAC National Accelerator Laboratory, Menlo Park, California 94025, USA}
\author{S.~Schmidt}\affiliation{Erlangen Centre for Astroparticle Physics (ECAP), Friedrich-Alexander University Erlangen-N\"urnberg, Erlangen 91058, Germany}
\author{A.~Schubert}\affiliation{Physics Department, Stanford University, Stanford, California 94305, USA}
\author{D.~Sinclair}\affiliation{Physics Department, Carleton University, Ottawa, Ontario K1S 5B6, Canada}\affiliation{TRIUMF, Vancouver, British Columbia V6T 2A3, Canada}
\author{V.~Stekhanov}\affiliation{Institute for Theoretical and Experimental Physics, Moscow, Russia}
\author{M.~Tarka}\affiliation{Department of Physics and Astronomy, Stony Brook University, SUNY, Stony Brook, New York 11794, USA}
\author{T.~Tolba}\affiliation{Institute of High Energy Physics, Beijing, China}
\author{R.~Tsang}\altaffiliation{Now at Pacific Northwest National Laboratory, Richland, Washington, USA}\affiliation{Department of Physics and Astronomy, University of Alabama, Tuscaloosa, Alabama 35487, USA}
\author{P.~Vogel}\affiliation{Kellogg Lab, Caltech, Pasadena, California 91125, USA}
\author{J.-L.~Vuilleumier}\affiliation{LHEP, Albert Einstein Center, University of Bern, Bern, Switzerland}
\author{M.~Wagenpfeil}\affiliation{Erlangen Centre for Astroparticle Physics (ECAP), Friedrich-Alexander University Erlangen-N\"urnberg, Erlangen 91058, Germany}
\author{A.~Waite}\affiliation{SLAC National Accelerator Laboratory, Menlo Park, California 94025, USA}
\author{T.~Walton}\affiliation{Physics Department, Colorado State University, Fort Collins, Colorado 80523, USA}
\author{M.~Weber}\affiliation{Physics Department, Stanford University, Stanford, California 94305, USA}
\author{L.J.~Wen}\affiliation{Institute of High Energy Physics, Beijing, China}
\author{U.~Wichoski}\affiliation{Department of Physics, Laurentian University, Sudbury, Ontario P3E 2C6, Canada}
\author{G.~Wrede}\affiliation{Erlangen Centre for Astroparticle Physics (ECAP), Friedrich-Alexander University Erlangen-N\"urnberg, Erlangen 91058, Germany}
\author{L.~Yang}\affiliation{Physics Department, University of Illinois, Urbana-Champaign, Illinois 61801, USA}
\author{Y.-R.~Yen}\affiliation{Department of Physics, Drexel University, Philadelphia, Pennsylvania 19104, USA}
\author{O.Ya.~Zeldovich}\affiliation{Institute for Theoretical and Experimental Physics, Moscow, Russia}
\author{J.~Zettlemoyer}\affiliation{Physics Department and CEEM, Indiana University, Bloomington, Indiana 47405, USA}
\author{T.~Ziegler}\affiliation{Erlangen Centre for Astroparticle Physics (ECAP), Friedrich-Alexander University Erlangen-N\"urnberg, Erlangen 91058, Germany}

\collaboration{EXO-200 Collaboration}

\date{\today}

\begin{abstract}

  Results from a search for neutrinoless double-beta decay (\nonubb) of
  $^{136}$Xe are presented using the first year of data taken with the upgraded
  EXO-200 detector.
  Relative to previous searches by EXO-200, the energy resolution of the
  detector has been improved to $\sigma/E$=1.23\%, the electric field in the
  drift region has been raised by 50\%, and a system to suppress radon in the
  volume between the cryostat and lead shielding has been implemented.  
  In addition, analysis
  techniques that improve topological discrimination between \nonubb and
  background events have been developed. Incorporating these hardware and
  analysis improvements, the median 90\% confidence level \nonubb half-life
  sensitivity after combining with the full data set acquired before the 
  upgrade has increased 2-fold to $3.7 \cdot 10^{25}$~yr. 
  No statistically significant evidence for \nonubb is observed, leading to a lower
  limit on the \nonubb half-life of $1.8\cdot10^{25}$~yr at the 90\% confidence
  level.

\end{abstract}


\maketitle

Neutrinoless double-beta decay (\nonubb), in which a nucleus with mass number
$A$ and charge $Z$ undergoes the decay $(A,Z)\rightarrow(A,Z+2)+2e^-$ with the
emission of no neutrinos~\cite{PhysRev.56.1184}, provides the most sensitive
test of the Majorana nature of neutrinos~\cite{DellOro:2016tmg}.  While
the corresponding two-neutrino double-beta decay (\twonubb) has been observed for
several nuclides~\cite{PDG16}, the observation of \nonubb
would provide direct evidence for a beyond-the-Standard-Model process that
violates lepton number conservation   
and constrain the absolute neutrino
mass scale~\cite{Engel:2016xgb}. 
Motivated by these implications, a
variety of experiments are searching for \nonubb in a number of nuclides,
reaching half-life sensitivities in excess of $10^{25}$
years~(e.g.~\cite{EXO_Nature,KamLAND-Zen:2016pfg,Agostini:2017iyd}), with
the most stringent for \xeiso at $5.6 \cdot 10^{25}$~yr~\cite{KamLAND-Zen:2016pfg}.

EXO-200 is searching for \nonubb in \xeiso
(see~\cite{EXO_Nature,Albert2013,Auger:2012gs} for a detailed description).  The
detector consists of a cylindrical time projection chamber (TPC) filled with liquid xenon (LXe) enriched to 80.6\% \xeiso.
The TPC is split into two drift regions by a common cathode, each with radius 
$\sim$18~cm and drift length $\sim$20~cm.
Energy
depositions in the LXe produce both scintillation light and ionization. 
The ionization charge is read out
after being drifted to crossed-wire planes at each anode by an electric
field, while the scintillation light is collected by arrays of avalanche photodiodes (APDs)~\cite{Neilson:2009kf}
located behind the wire planes. For each interaction, the location of the deposited charge in the
directions perpendicular to the drift field ($x$ and $y$) is determined from the
wire signals. The $z$ position is reconstructed from the time delay
between the prompt light signal and the delayed charge signals, using the
measured ionization drift velocity~\cite{EXO_diffusion}. 
The total energy deposited is determined from
the combination of the charge and light signals, optimally accounting for their 
anticorrelation~\cite{Conti2003}.

The LXe is housed in a thin-walled copper vessel, and surrounded by several
layers of passive and active shielding, including $\sim$50 cm of HFE-7000 cryofluid~\cite{3m} and
$\sim$25~cm of lead in all directions~\cite{Auger:2012gs}. A plastic
scintillator muon veto surrounds the experiment on four
sides~\cite{Albert2013,EXO_cosmogenics}.  The detector is located at the Waste
Isolation Pilot Plant (WIPP) near Carlsbad New Mexico, which provides an
overburden of 1624$^{+22}_{-21}$ meters water equivalent~\cite{EXO_cosmogenics}.

\begin{figure}[t]
\centering
\includegraphics[width=\columnwidth]{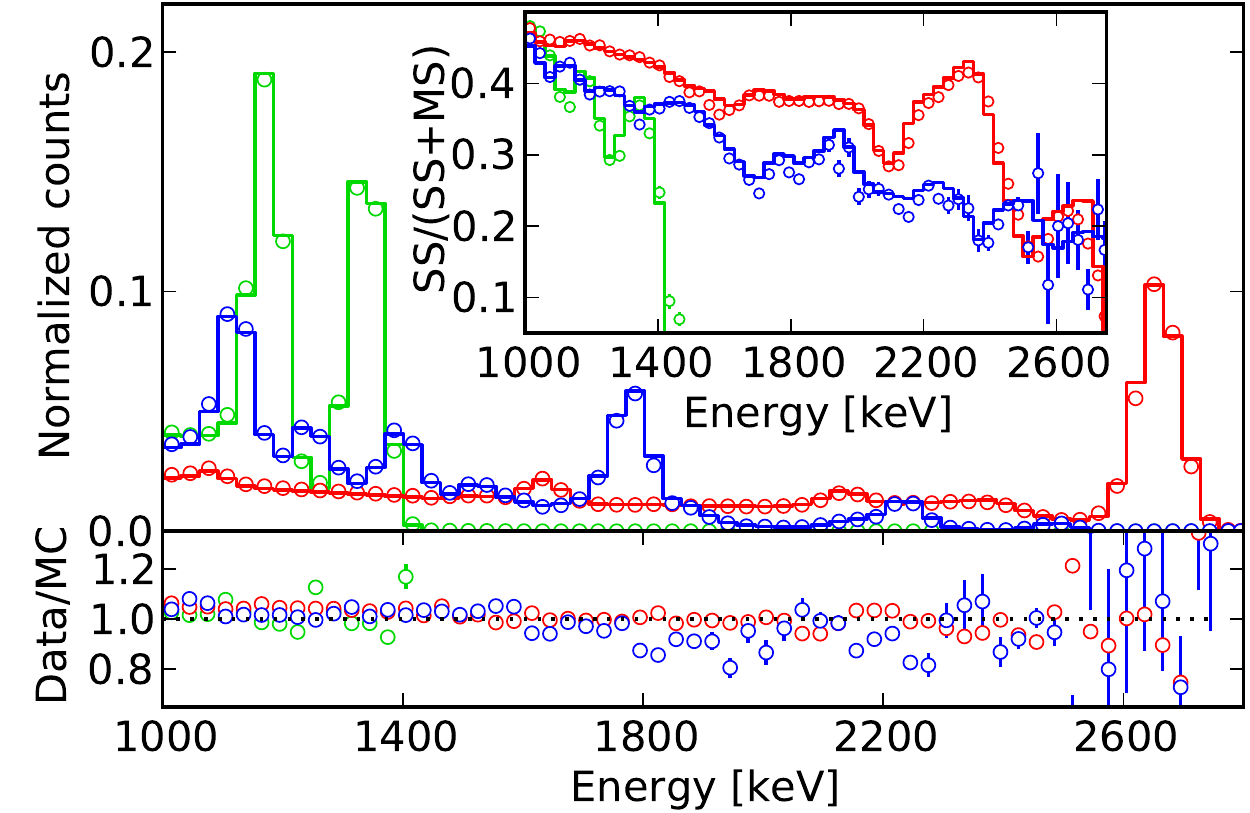}
\caption{(Color online) Comparison between SS events in \PII data (open markers) and
  MC (lines) for calibrations using \cosrc (green), \rasrc (blue), and \thsrc
  (red) sources positioned near the cathode.  
  The bottom shows the ratio between data and MC. 
  The inset compares the
  corresponding SS fraction, SS/(SS+MS), for the calibration data and MC.}
\label{fig:energy_source_agree}
\end{figure}

To reconstruct events in the TPC, charge and light signals are first grouped
into individual energy deposits within each event.  Events with a single
reconstructed deposit are identified as ``single-site'' (SS), while events with
multiple, spatially-separated deposits are denoted as ``multi-site'' (MS).  
This topological SS/MS classification 
has been used in previous EXO-200 analyses and provides  
discrimination between $\gamma$ backgrounds, which are 
primarily MS, and the \nonubb signal, which is primarily SS.  A
detailed detector Monte Carlo (MC) simulation based on Geant4~\cite{GEANT42006}
is used to model the energy deposits produced in the LXe by various backgrounds
and the \nonubb signal. The MC simulation propagates the charge deposits through
the detector and produces simulated waveforms for each readout channel and
event. The MC waveforms are processed using the same analysis framework as the data
waveforms. 
In order to calibrate the detector energy scale and validate the accuracy of the
MC simulation, runs are taken using $\gamma$ sources positioned
within 10~cm of the LXe vessel at locations around the cathode plane and at both
anode planes.  Figure~\ref{fig:energy_source_agree} shows the agreement between
the MC simulation and data acquired with \cosrc, \rasrc, and \thsrc
sources. 

EXO-200 has previously reported results on a search for \nonubb~\cite{EXO_Nature}
using 80\% of the data from its first run (``\PI''), which spans from
Sept.~2011 to Feb.~2014.
In Feb.~2014, EXO-200 was forced to suspend operations, because of
accidents at the WIPP facility and recover the Xe from the detector.
After access to the experiment was regained in early 2015, the detector was
recommissioned and refilled with LXe in Jan. 2016.  Between
Jan. and May~2016 the
detector was upgraded with
new electronics primarily aimed at improving the APD read-out noise. 
In addition, a system 
was installed to suppress \Rn in the air
gap between the copper cryostat and the lead shield. 
After installation of this system, direct sampling
of the air
indicated that the average \Rn level was reduced by more than a factor of 10. 
Fits to the energy and location of backgrounds in physics data also indicated a lower best-fit 
value for this background component, although more data are required 
to demonstrate a statistically significant reduction.
Finally, the
electric field in the drift region of the detector was raised from 380~V/cm 
(cathode voltage, $V_C=-8$~kV) 
to 567~V/cm ($V_C=-12$~kV). 
The data taking run with the upgraded detector
began in May~2016 (``\PII'').

\begin{figure}[t]
\centering
\includegraphics[width=\columnwidth]{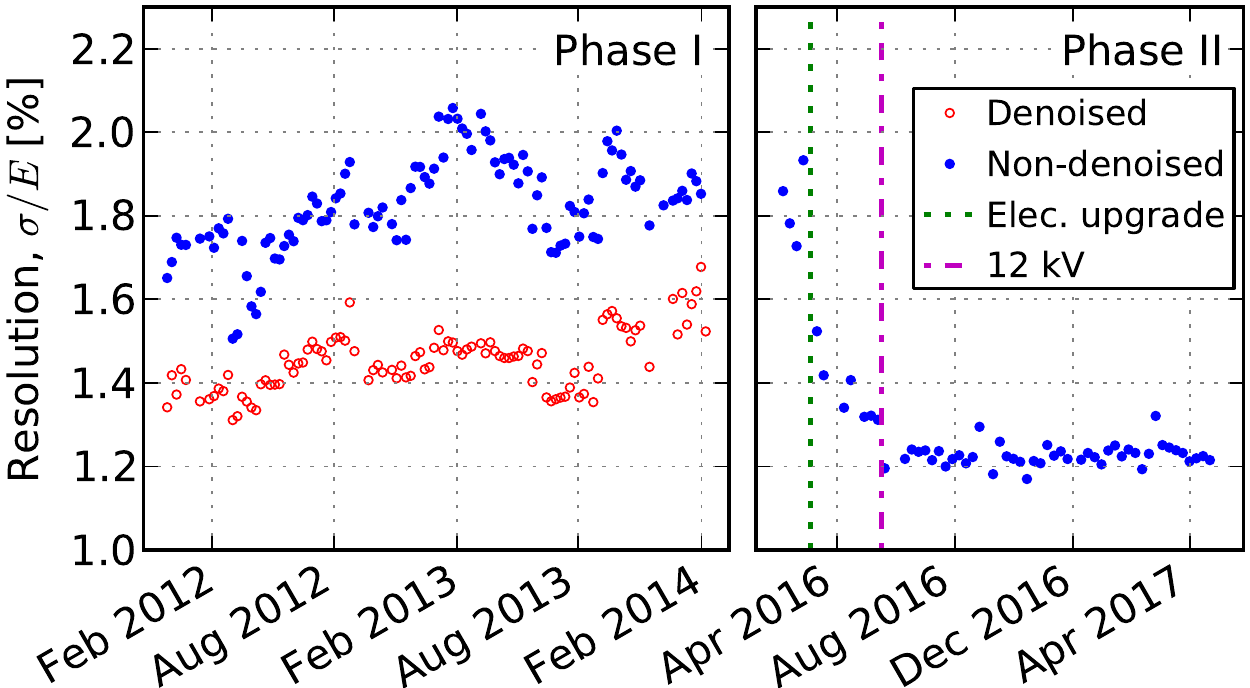}
\caption{(Color online) Measured energy resolution, $\sigma/E$, for the 2615~keV
  $^{208}$Tl $\gamma$ line in calibration data taken at the cathode position
  throughout \PI and \PII.  The measured resolution before (blue) and after
  (red) applying the software de-noising algorithm in \PI are shown. 
  The data acquired between restart of operations and the start of \PII, 
  when $V_C$ was raised to -12~kV, were not used in the current analysis.
}
\label{fig:resolution}
\end{figure}

The primary goal of the electronics upgrade was to minimize 
the APD read-out noise observed 
in \PI.  While this noise
was accounted for in previous analyses and partially suppressed using a software ``de-noising''
algorithm~\cite{EXO_denoising}, the hardware upgrade provides substantially 
improved performance.  The effect on the energy resolution is shown in
Fig.~\ref{fig:resolution}.  In \PI, the SS resolution at the \nonubb
decay energy of $Q_{\beta\!\beta} = 2457.83\pm0.37$~keV~\cite{Redshaw:2007} after
applying the software de-noising algorithm is
$\sigma/E(Q_{\beta\!\beta})=1.38$\%, averaged over live time and position.
In \PII, this figure is 1.23\% and its time variation is greatly reduced.
These values account for the spatial variation of the resolution,  
including events taken with the calibration source behind the anodes. 
Because of the source's proximity to the readout when at the anode, 
these events present better energy resolution than those in Fig.~\ref{fig:resolution}.

The selection cuts for this analysis closely follow those used in previous
EXO-200 analyses~\cite{EXO_Nature}. 
Both \PII data and the previously examined \PI data were ``blinded'' to remove 
candidate \nonubb 
events in the energy region between 2345~keV and 2570~keV.
After data quality cuts~\cite{Albert2013}, the total 
exposure considered here is 596.7~d and 271.8~d for \PI and \PII, respectively.  

Only a
fiducial volume (FV) within the detector is considered. The FV selection requires the
position of all charge deposits in an event to be reconstructed within a hexagon
with apothem of 162~mm and
more than $10$~mm away from the anode and cathode wire planes, as well as
from the cylindrical PTFE reflector inside the field-shaping rings.  
This FV corresponds to 74.7~kg
of $^{136}$Xe, i.e. $3.31\cdot10^{26}$~atoms, 
resulting in a total exposure of 
177.6~kg$\cdot$yr or 1307~mol$\cdot$yr. 
The individual exposure in \PI and \PII
are 122~kg$\cdot$yr and 55.6~kg$\cdot$yr, respectively, 
or 898~mol$\cdot$yr and 409~mol$\cdot$yr.

To suppress backgrounds correlated in time,
events are required
to have only a single reconstructed scintillation signal and to occur $>1$~s
from all other reconstructed events.
The corresponding \nonubb signal reconstruction efficiency
is found to be consistent between phases within errors, 
$82.4 \pm 3.0$\% ($80.8 \pm 2.9$\%) for \PI (\PII). The
inefficiency is dominated by the 1~s anticoincidence cut and by incomplete
reconstruction of \nonubb events with small, separated energy deposits from
bremsstrahlung.
Its errors are
determined from the
difference in the observed absolute rate for calibration source data and MC
using the known source activity, and measurements of the individual cut
efficiencies for low-background \twonubb events. 
It includes the estimation of the uncertainty in the FV, 2.8\% in \PI 
(2.6\% in \PII), the dominant term in this error.

This analysis introduces a cut to
reduce the rate of background events arising from cosmogenically produced
$^{137}$Xe~\cite{EXO_cosmogenics}, which decays via $\beta$ emission with a
total energy of 4173~keV~\cite{Xe137Qvalue}.  Events in coincidence with the 
muon veto detector, and depositing 
energy consistent with the cascade $\gamma$s 
emitted after the neutron capture on $^{136}$Xe, are used to veto 
subsequent events in the same TPC half within 19.1~min, 
corresponding to  $5 \cdot T^{^{137}\text{Xe}}_{1/2}$.
This cut was estimated to reduce the number of $^{137}$Xe events 
by $23\pm8$\%, with a loss in exposure of 3.5\% (2.8\%)
in \PI (\PII). This reduction is consistent with the $^{137}$Xe rate 
entirely attributed to cosmogenic sources~\cite{EXO_cosmogenics}.

\begin{figure*}[t]
\centering
\includegraphics[width=\linewidth]{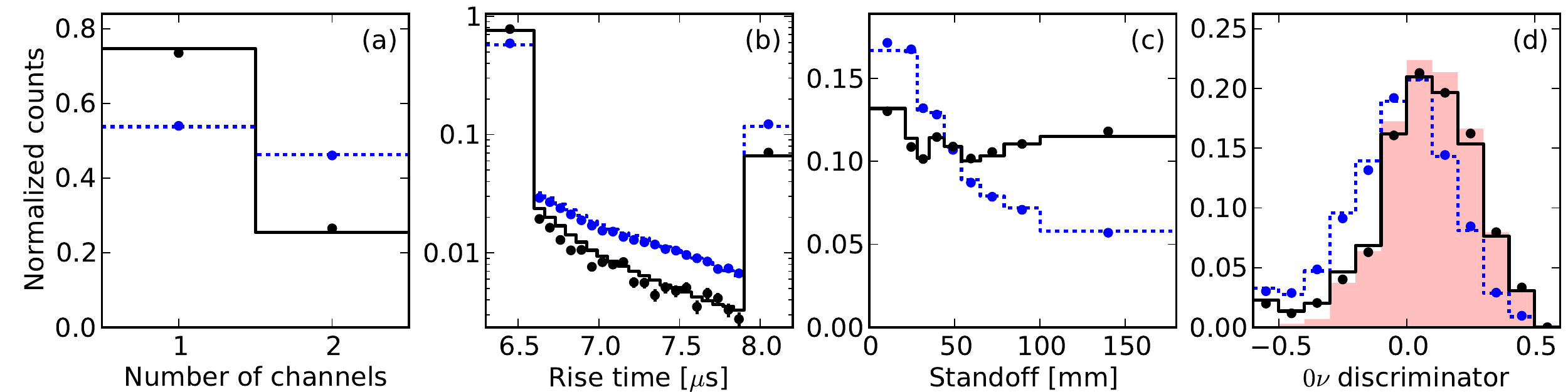}
\caption{(Color online) Comparison between data (dots) and MC (solid/dashed
  lines) for the individual variables used in the BDT and the overall
  discriminator distribution.  Both source calibration data using the \rasrc
  source at the cathode (blue dashed) and the background-subtracted \twonubb spectrum
  from low background data (black solid) are shown.  
  Only SS events are depicted in the plots.
  Statistical error bars on the data
  points are included, but are typically smaller than the marker size.  The
  expected BDT discriminator distribution for a \nonubb signal from MC is indicated by the red
  filled region. All distributions are normalized by the area, and the edge bins account 
  for overflow.}
\label{fig:bdt_source_agree}
\end{figure*}

New techniques have been developed to further improve 
$\gamma$-background rejection among events classified as SS by 
using the detailed topological
information available for each interaction in the TPC.  By implementing
transverse electron diffusion (coefficient $D_t =
  55\ \mathrm{cm}^2$/s~\cite{EXO_diffusion}) and the 
three-dimensional geometry of the wire planes in the detector model, 
the number of channels that collect charge signals  
(denoted as ``number of channels'') is now
accurately simulated.  Figure~\ref{fig:bdt_source_agree}~(a) shows that SS
$\gamma$ backgrounds are more likely to deposit energy on more than one
neighboring channel than the $\beta\!\beta$ signal.  In addition, 
extending this concept to the $z$-direction, the
distribution of the rise time of the charge pulse (defined as the time between
collection of 5\% to 95\% of the total signal) is more likely to
extend to large values for $\gamma$ backgrounds relative to $\beta\!\beta$
events (Fig.~\ref{fig:bdt_source_agree}~(b)).  Finally,  
the ``standoff distance,'' denoting the minimum distance
between a cluster and the closest TPC surface, excluding the cathode, is used to
constrain backgrounds originating from sources external to the LXe (Fig.~\ref{fig:bdt_source_agree}~(c)).

A multivariate discriminator was developed by combining these 
topological variables
in a boosted decision tree (BDT) using the TMVA software package~\cite{TMVA}.
The separation between SS
\nonubb and the most prominent $\gamma$ backgrounds 
($^{238}$U, $^{232}$Th, and $^{60}$Co) was maximized 
using a subset of the MC. Its performance
was then tested on a statistically independent MC data set.
Agreement between data and MC for calibration sources for both the BDT and its constituent variables was used to validate its performance for
the main backgrounds with high statistics, while the corresponding distributions
for signal-like events were investigated using a pure sample of 
\twonubb SS events with energy near the $Q_{\beta\!\beta}$.
The ranked importance of the individual discriminator
variables\textemdash defined as the weighted fraction of decision tree cuts for
which each variable was used\textemdash was found to be
42\%, 39\%, and 19\% for the rise time, 
standoff distance, and number of channels, respectively.

Figure~\ref{fig:bdt_source_agree} shows a comparison between the simulated and
observed data distributions for calibration sources, and for the
measured background-subtracted \twonubb distribution.
Overall,
the data and MC distributions for the input variables and the overall discriminator
agree to better than 10\% at every bin.  The detailed binning and range used for each
variable was optimized to minimize systematic errors arising from imperfections
in the MC simulation, while maintaining as much discriminating power as
possible.  As described below, the systematic errors resulting from the
differences between the data and MC distributions are evaluated using toy MC
studies. These residual differences contribute a sub-dominant uncertainty to the backgrounds and signal
efficiency.

To search for a \nonubb signal, the \PI and \PII data 
are separately fit to models using a binned maximum-likelihood (ML) fit. 
These models consist of the \nonubb signal and backgrounds 
originating from the detector and surrounding materials. 
The background model closely follows that used in previous EXO-200
analyses~\cite{EXO_Nature}.
The shape of the spectrum for each of the fit observables is determined from the
MC simulation for each background and signal component.  The energy spectra for
the SS and MS data are fit simultaneously,
and unlike the previous analysis of \PI data~\cite{EXO_Nature}, 
the BDT variable  (including the standoff variable) 
is added as a fit dimension for the SS data. 
Toy studies indicated that the addition of the BDT or standoff to the MS fit did not
enhance sensitivity for this search.
Systematic errors are 
included in the ML fit as nuisance parameters, 
constrained by normal distributions.  
An overall normalization constrained to unity is 
included to account for the error on the detection efficiency.

The balance between SS and MS events, parameterized by the ``SS fraction,'' 
is allowed to vary around the expected value from MC for each component
within a systematic error. 
This error was determined by comparing
the SS fraction for source calibration data and MC, as shown in
the inset to Fig.~\ref{fig:energy_source_agree}.
Averaging over all calibration positions acquired throughout \PI (\PII) 
gives a relative error on the SS fraction of 5.0\% (8.8\%).  An
85\% correlation between the SS fractions of the $\gamma$-like components 
is included in the constraint, justified by similar levels observed in 
calibration source data. 

\begin{figure*}[th!]
\centering
\includegraphics[width=\linewidth]{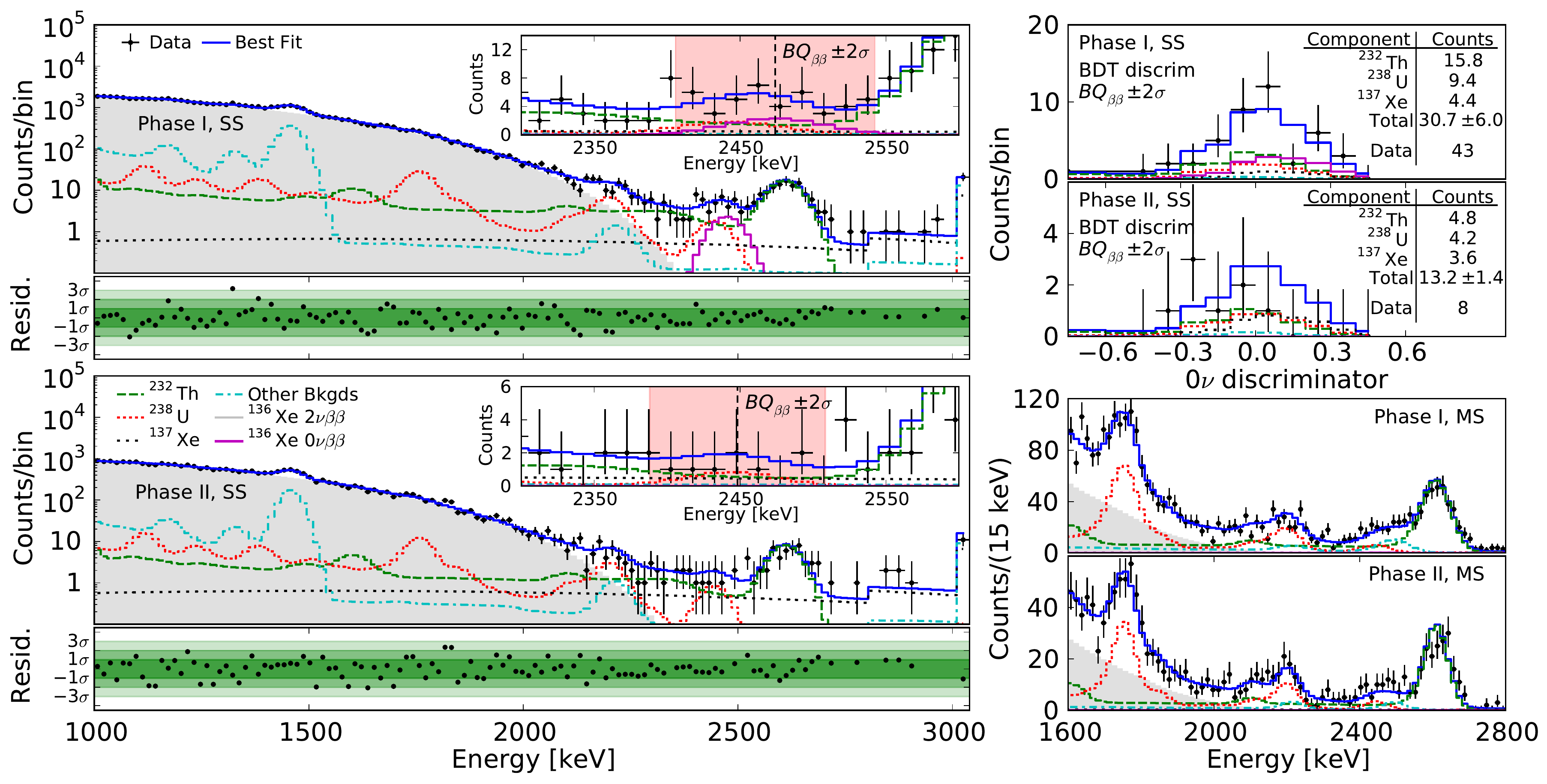}
\caption{(Color online) Best fit to the low background data SS energy spectrum
  for \PI (top left) and \PII (bottom left). The energy bins are 15~keV and 30~keV 
  below and above 2800~keV, respectively.
  The inset shows a zoomed in view 
  around the best-fit value for $B Q_{\beta\!\beta}$.  
  (top right) Projection of events  within $B Q_{\beta\!\beta}\pm2\sigma$ 
  on the BDT fit dimension. (bottom right) MS energy spectra above the 
  \isot{K}{40} $\gamma$-line.}
\label{fig:energy_fit}
\end{figure*}

Since the ML fit relies on accurately modeling the shapes of the various
background components, the impact of shape differences 
between data and MC was investigated for each fit observable 
(see Figs.~\ref{fig:energy_source_agree} and~\ref{fig:bdt_source_agree}).
In these studies, the shapes of the $\gamma$-originated 
background components are corrected 
by using the residual differences
between calibration source data and simulation, while
the shapes of the SS $\beta$ components are corrected by 
using the residual differences of the measured background-subtracted \twonubb spectrum. 
A large number of
simulated data sets were drawn from the best-fit background model using the
corrected PDFs, and were fit with the original simulated shapes.  The resulting bias
between the fitted and true value of backgrounds near $Q_{\beta\!\beta}$ 
is included as an additional systematic error on the normalization
of the background components. Toy studies indicate that these
shape errors are 2.1\% (1.7\%) for \PI (\PII). 
The contribution to this error caused by spatial and
temporal energy resolution variations that are not fully accounted for
by the MC simulation was determined to be 1.5\% (1.2\%) in \PI (\PII).

U, Th, and Co background components simulated at locations 
different from the default ones were individually inserted into
the fit, and the resulting variation in the number of expected events 
near $Q_{\beta\!\beta}$ was determined.
These studies estimate the error due to uncertainty in the location of the
background model components to be 5.6\% (5.9\%) in \PI (\PII). 
All sources of systematic uncertainty on the background model near $Q_{\beta\!\beta}$ are treated as
uncorrelated and result in a 
total error of 6.2\% for both \PI and \PII, as summarized in Tab.~\ref{tab:shape-syst}.

\begin{table}[h]
\caption{Systematic errors on the determination of the number of events near $Q_{\beta\!\beta}$.}
\label{tab:shape-syst}
\begin{tabular}{lcc}
\hline
\hline
Source & \PI & \PII \\
\hline
Signal detection efficiency & 3.0\% & 2.9\% \\
\hline
Background errors &  &  \\
\-\hspace{0.5cm} Spectral shape agreement & 2.1\% & 1.7\% \\
\-\hspace{0.5cm} Background model & 5.6\% & 5.9\% \\
\-\hspace{0.5cm} Energy scale and resolution & 1.5\% & 1.2\% \\ \cline{2-3}
Total & 6.2\% & 6.2\% \\
\hline 
\hline 
\end{tabular}
\end{table}

Two final constraints on the measured \Rn concentration in the LXe and relative
rate of cosmogenically produced backgrounds were included in the fit, but
verified to be unchanged from previous analyses~\cite{EXO_Nature} for both \PI and \PII.

The analysis further accounts for a possible
difference in the reconstructed energy for $\beta$-like events, $E_\beta$, relative to the
energy scale determined from the $\gamma$ calibration sources, $E_\gamma$. 
This difference is expressed through a multiplicative 
constant, $B$, that scales the energy for all $\beta$-like
components, $E_\beta = B E_\gamma$, which is allowed to float freely in the fit.  
$B$ is highly constrained by the \twonubb spectrum, and 
consistent with unity to the sub-percent level. 

After ``unblinding'' the combined data set, no statistically significant
evidence for \nonubb was observed. 
A lower limit on the
half-life of $T_{1/2}>1.8\cdot10^{25}$~yr at the 90\% confidence level (CL) was
derived from the ML fits after profiling over nuisance parameters. 
The data from each phase is fit separately and the profiles added 
together considering the difference in live time and signal detection efficiency. 
No correlation was considered between these two profiles. This conservative 
assumption was estimated to negligibly change the expected sensitivity.
The profile-likelihood distribution was determined from toy MC simulations, 
following the same procedure to combine phases, and 
found to be in good agreement with Wilks's theorem~\cite{wilks1938,cowan1998statistical}. 
Under the assumption that neutrinos are Majorana particles,
this corresponds to an upper limit on the Majorana neutrino mass,
$\langle m_{\beta\!\beta} \rangle < (147-398)$~meV~\cite{DellOro:2016tmg},
using the nuclear matrix elements of~\cite{ibm2:2015,Rodr:2010,QRPA:2014,Menendez2009139,SkyrmeQRPA:2013} 
and phase space factor from~\cite{Kotila:2012zza}.
The best-fit value for the \nonubb component is consistent with the null hypothesis
at 1.5$\sigma$, corresponding to a $p$-value of 0.12.

The results of the ML fits are presented in Fig~\ref{fig:energy_fit}. 
The measured \twonubb rates were found to be
consistent with~\cite{Albert2013}. 
The best-fit contributions from the primary background components within 
$B Q_{\beta\!\beta}\pm2\sigma$ are summarized in the inset table in 
Fig~\ref{fig:energy_fit} (top right).
The best-fit total event rate is $(1.5\pm0.3)\cdot10^{-3}$~kg$^{-1}$yr$^{-1}$keV$^{-1}$  
[$(1.6\pm0.2)\cdot10^{-3}$~kg$^{-1}$yr$^{-1}$keV$^{-1}$] 
when normalized to the full mass including all Xe isotopes
for \PI [Phase~II]. 

The median 90\% CL
sensitivity was estimated from toy MC studies 
to be $3.7\cdot 10^{25}$~yr.  This represents a
factor of $\sim$2 improvement over the previous EXO-200
search~\cite{EXO_Nature}.  
In comparison to fits using the energy spectra and SS/MS classification alone,
or with the addition of only the standoff distance, the use of the BDT
discriminator provides a $\sim$15\% increase in sensitivity.  

The individual \PI and \PII data set lower limits
of $1.0\cdot10^{25}$~yr and $4.4\cdot10^{25}$~yr at the 90\% CL,
respectively, with corresponding median sensitivity of $2.9\cdot 10^{25}$~yr and $1.7\cdot 10^{25}$~yr.  
Because of the detector upgrades and improved topological discrimination
described here, the \PII sensitivity from this
analysis is already comparable to that of the previous EXO-200 \nonubb
search~\cite{EXO_Nature} with an exposure that is half the size.  
The combined analysis of the \PI and
\PII data provides one of the most sensitive searches for \nonubb for
any isotope~\cite{KamLAND-Zen:2016pfg,Agostini:2017iyd} to date.  
Further operation of the upgraded
detector is expected to continue improving sensitivity to \nonubb, and 
holds promise for nEXO~\cite{nEXO-sens}, 
a tonne-scale LXe TPC being designed to reach half-life sensitivity of $\sim10^{28}$~yr.

\begin{acknowledgments}
EXO-200 is supported by DOE and NSF in the United States, NSERC in
Canada, SNF in Switzerland, IBS in Korea, RFBR in Russia, DFG in
Germany, and CAS and ISTCP in China. EXO-200 data analysis and
simulation uses resources of the National Energy Research Scientific
Computing Center (NERSC).  We gratefully acknowledge the KARMEN
collaboration for supplying the cosmic-ray veto detectors, and the
WIPP for their hospitality.
\end{acknowledgments}

\bibliography{exo_PRL_0nubb}

\end{document}